\pdfoutput=1
\documentclass[aps,reprint,floatfix,superscriptaddress]{revtex4-1} 

\usepackage{amsmath,amsfonts,amssymb} %
\usepackage{mathtools} %
\usepackage{wasysym} %
\usepackage{bm} %
\usepackage[bbgreekl]{mathbbol} %
\usepackage{graphicx} %
\usepackage[dvipsnames,table]{xcolor} %
\usepackage{tabularx} %
\usepackage[acronym]{glossaries} %
\usepackage{braket} %
\usepackage[binary-units=true,detect-weight=true]{siunitx} %
\usepackage{fancyhdr} %
\usepackage{microtype} %
\usepackage{natmove}
\usepackage[caption=false]{subfig}
\usepackage[byname]{smartref} %
\usepackage[breaklinks, %
colorlinks, linkcolor=Blue, citecolor=Blue, urlcolor=Blue]{hyperref} %
\usepackage[version=3]{mhchem}
\usepackage{wrapfig}
\AtBeginDocument{\usepackage{booktabs}} 

\DeclareSIUnit[number-unit-product = {\,}]\cal{cal}
\def\MyTitle{A posteriori corrections to the Iterative Qubit Coupled
  Cluster method to minimize the use of quantum resources in
  large-scale calculations} %
\def\MyAuthora{Ilya G. Ryabinkin} %
\def\MyAuthorb{Artur F. Izmaylov} %
\def\MyAuthorc{Scott N. Genin} %

\def\MySubject{Quantum computing, quantum chemistry} %

\hypersetup{%
  pdftitle={\MyTitle{}}, %
  pdfauthor={\MyAuthora{}, \MyAuthorb{}, and \MyAuthorc{}}, %
  pdfsubject={\MySubject{}}, %
  pdfkeywords={quantum computing, NISQ devices, electronic structure
    methods} %
}

\graphicspath{{pics/}}

\newcolumntype{Y}{>{\centering\arraybackslash}X}

\newacronym[longplural={degrees of freedom}, firstplural={degrees of
  freedom (DOF)}, plural={DOF}]{DOF}{DOF}{degree of freedom} %
\newacronym[longplural={equations of motion}, firstplural={equations
  of motion (EOM)}, plural={EOM}]{EOM}{EOM}{equation of motion} %

\newacronym{NISQ}{NISQ}{noisy intermediate-scale quantum}

\newacronym{JW}{JW}{Jordan--Wigner} %
\newacronym{BK}{BK}{Bravyi--Kitaev} %

\newacronym{QPE}{QPE}{quantum phase estimation} %
\newacronym{VQE}{VQE}{variational quantum eigensolver} %
\newacronym{QMF}{QMF}{qubit mean-field} %
\newacronym{QCC}{QCC}{qubit coupled cluster} %
\newacronym{iQCC}{iQCC}{iterative qubit coupled cluster} %
\newacronym{PQA}{PQA}{parametrized quantum annealing} %
\newacronym{DIS}{DIS}{direct interaction set} %

\newacronym{CAS}{CAS}{complete active space} %
\newacronym{PES}{PES}{potential energy surface} %
\newacronym{PEC}{PEC}{potential energy curve} %
\newacronym{MO}{MO}{molecular orbital} %

\newacronym{CI}{CI}{configuration interaction} %
\newacronym{FCI}{FCI}{full configurational interaction} %
\newacronym{CASCI}{CASCI}{complete active space configurational
  interaction} %
\newacronym{MCSCF}{MCSCF}{multiconfigurational self-consistent
  field} %
\newacronym{CASSCF}{CASSCF}{complete active space self-consistent
  field} %
\newacronym{CC}{CC}{coupled cluster} %
\newacronym{UCC}{UCC}{unitary coupled cluster} %
\newacronym{UCCSD}{UCCSD}{unitary coupled cluster singles and
  doubles} %
\newacronym{CCSD}{CCSD}{coupled-cluster singles and doubles} %
\newacronym{CCSD-T}{CCSD(T)}{coupled-cluster singles and doubles and
  non-iterative triples} %
\newacronym{RHF}{RHF}{restricted Hartree--Fock} %
\newacronym{UHF}{UHF}{unrestricted Hartree--Fock} %
\newacronym{DMRG}{DMRG}{density-matrix renormalization group} %
\newacronym{DFT}{DFT}{density-functional theory} %
\newacronym{ENPT}{ENPT}{Epstein-Nesbet perturbation theory} %
\newacronym{MP}{MP}{M{\o}ller--Plesset perturbation theory} %
\newacronym{MRMP2}{MRMP2}{second-order multi-reference M{\o}ller--Plesset perturbation theory} %

\newacronym{SQP}{SQP}{sequential quadratic programming} %
\newacronym{MMA}{MMA}{method of moving asymptotes} %

\newcommand{\E}{\textrm{e}} %
\newcommand{\I}{\mathrm{i}\mkern1mu} %
\def\be{\begin{equation}} %
\def\ee{\end{equation}} %
\def\bea{\begin{eqnarray}} %
\def\eea{\end{eqnarray}} %

\begin{document}

\title{\MyTitle}

\author{\MyAuthora{}} %
\email{ilya.ryabinkin@otilumionics.com} %
\affiliation{OTI Lumionics Inc., 100 College St. \#351, Toronto,
  Ontario\, M5G 1L5, Canada} %

\author{\MyAuthorb{}} %
\email{artur.izmaylov@utoronto.ca} \affiliation{Department of Physical
  and Environmental Sciences, University of Toronto Scarborough,
  Toronto, Ontario, M1C 1A4, Canada; and Chemical Physics Theory
  Group, Department of Chemistry, University of Toronto, Toronto,
  Ontario, M5S 3H6, Canada}

\author{\MyAuthorc{}} %
\email{scott.genin@otilumionics.com} \affiliation{OTI Lumionics Inc.,
  100 College St. \#351, Toronto, Ontario\, M5G 1L5, Canada} %

\date{\today}

\begin{abstract}
  The \gls{iQCC} method is a systematic variational approach to solve
  the electronic structure problem on universal quantum computers. It
  is able to use arbitrarily shallow quantum circuits at expense of
  iterative canonical transformation of the Hamiltonian and rebuilding
  a circuit. Here we present a variety of \emph{a posteriori}
  corrections to the \gls{iQCC} energies to reduce the number of
  iterations to achieve the desired accuracy. Our energy corrections
  are based on a low-order perturbation theory series that can be
  efficiently evaluated on a classical computer. Moreover, capturing a
  part of the total energy perturbatively, allows us to formulate the
  qubit active-space concept, in which only a subset of all qubits is
  treated variationally. As a result, further reduction of quantum
  resource requirements is achieved. We demonstrate the utility and
  efficiency of our approach numerically on the examples of 10-qubit
  \ce{N2} molecule dissociation, the 24-qubit \ce{H2O} symmetric
  stretch, and 56-qubit singlet-triplet gap calculations for the
  technologically important complex,
  tris-(2-phenylpyridine)iridium(III) \ce{Ir(ppy)3}.
\end{abstract}

\glsresetall

\maketitle

\section{Introduction}

Electronic structure calculations~\cite{Helgaker:2000} on
universal-gate \gls{NISQ}~\cite{Preskill:2018/quant/79} devices are
challenging because of the limited number of qubits, limited
connectivity, short coherence times, and noisy measurements. In light
of these limitations, algorithms based on the \gls{VQE}
framework~\cite{Peruzzo:2014/ncomm/4213, Wecker:2015/pra/042303}, a
hybrid quantum-classical variational scheme, show the most promise.
Central to \gls{VQE} is a unitary ${\hat U}(\boldsymbol \tau)$, where
$\boldsymbol \tau$ is a vector of numerical parameters, acting on an
initial state of a quantum register (an initial wavefunction)
$\ket{0}$. Once ${\hat U}(\boldsymbol\tau)$ is fixed, the variational
energy estimate
\begin{equation}
  \label{eq:VQE_energy_func}
  E(\boldsymbol\tau) = \braket{0|{\hat U}^\dagger(\boldsymbol\tau) \hat
    H {\hat U}(\boldsymbol\tau) |0},
\end{equation}
is evaluated on a quantum computer by measuring terms (or groups of
terms)~\cite{Izmaylov:2019/cs/3746, Zhao:2019/arXiv/1908.08067,
  Huggins:2019/arXiv/1907.13117, Knill:2007/pra/012328,
  Izmaylov:2020/jctc/190, Yen:2020/jctc/2400,
  Verteletskyi:2020/jcp/124114, Yen:2020/arXiv/2007.01234} of the
qubit Hamiltonian
\begin{equation}
  \label{eq:qubitH}
  \hat H = \sum_k C_k \hat P_k,
\end{equation}
where $C_k$ are numerical coefficients determined by the electronic
Hamiltonian of a problem, and
\begin{equation}
  \label{eq:pauli_word_def}
  \hat P_k = \prod_{i} {\hat\sigma_i^{(k)}}
\end{equation}
are strings of Pauli elementary
$\hat\sigma_i \in \{\hat x_i, \hat y_i, \hat z_i\}$ operators acting
on the $i^\text{th}$, $i = 1 \dots n$, qubit~\cite{Nielsen:2010}. We
call these strings ``Pauli words'' for brevity. The values of
$E(\boldsymbol\tau)$ are used by a classical computer to update
amplitudes $\boldsymbol\tau$ in the direction that lowers the energy.
The processes is repeated until the variational minimum of energy is
obtained.

The ansatz for ${\hat U}(\boldsymbol \tau)$ must satisfy certain
requirements. First, it must be either directly represented as a
sequence of quantum gates or readily converted to it. Secondly, it has
to be accurate already for the small number of variational parameters
and demonstrate rapid convergence with increasing the number of them.
Finally, it must be systematically improvable. Several forms of
${\hat U}(\boldsymbol\tau)$ were explored in literature. The
``hardware-efficient'' form~\cite{Kandala:2017/nature/242,
  Kandala:2019/nature/491} encodes $\hat U(\boldsymbol\tau)$ directly
as a set of gates available on a particular hardware, but by far the
most popular approach is based on the \gls{UCC} ansatz and its
generalizations~\cite{Peruzzo:2014/ncomm/4213,
  Mcclean:2016/njp/023023, OMalley:2016/prx/031007,
  Romero:2018/qct/014008, Hempel:2018/prx/031022,
  Nam:2019/ArXiv/1902.10171, Lee:2019/jctc/311},
\begin{align}
  \label{eq:UCC}
  \hat U(\boldsymbol \tau) & = \E^{\hat T(\boldsymbol \tau) - \hat
                             T^\dagger(\boldsymbol \tau)}
\end{align}
where $\hat T$ and $\hat T^\dagger$ are sums of coupled-cluster
$K$-fold fermion excitation $\hat T_K$ and de-excitation
$\hat T_K^\dagger$ operators, respectively. To be used in \gls{VQE},
the \gls{UCC} ansatz must be first converted to a product form,
\begin{equation}
  \label{eq:qcc_form}
  \hat U(\boldsymbol\tau) = \prod_j \exp(-\I\tau_j\hat P_j/2)
\end{equation}
in which the generators, Pauli words $P_j$, are inferred from
$\hat T_K$ and $\hat T_K^\dagger$. The difficulty here is that
$\hat T_K$ and $\hat T_K^\dagger$ are non-commuting so that after the
transformation to the qubit space, the ansatz ~\eqref{eq:qcc_form} is
order-dependent~\cite{Evangelista:2019/jcp/244112,
  Izmaylov:2020/arXiv/2003.07351}.

Recently, the \gls{VQE}-based methods that started with the
form~\eqref{eq:qcc_form} directly were
proposed~\cite{Ryabinkin:2018/jctc/6317, Grimsley:2018/nc/3007,
  Ryabinkin:2020/jctc/1055}. The ADAPT-VQE
method~\cite{Grimsley:2018/nc/3007} uses the set of \gls{UCCSD}
fermionic excitation operators converted to the qubit representation
as a pool of generators supplemented with a gradient-based ranking
procedure to select the ``most important'' ones. The \gls{iQCC}
method~\cite{Ryabinkin:2020/jctc/1055} gives up on the fermionic
excitation picture completely and introduces the concept of the
\gls{DIS}---a set of all possible operators that \emph{guarantees} the
first-order energy lowering being included in the
ansatz~\eqref{eq:qcc_form}. The \gls{DIS} can be efficiently
constructed on a classical computer given the qubit form of the
Hamiltonian~\eqref{eq:qubitH}.

The main challenge for these methods is finding generators among
$4^n-1$ possible that provide fast and systematic convergence. First
of all, none of the schemes that use polynomially-large spaces of the
fixed-rank fermionic excitations as a source of generators in
Eq.~\eqref{eq:qcc_form} can be exact, unless the same operators are
allowed to appear more than once~\cite{Grimsley:2018/nc/3007,
  Evangelista:2019/jcp/244112}. The repeated sampling is a strategy
adopted in ADAPT-VQE. Secondly, a simple random sampling of generators
is highly inefficient due to the so-called barren
plateaus~\cite{Mcclean:2018/nc/4812}. The \gls{iQCC} method uses the
iterative approach: Instead of a single-step construction and
optimization of a lengthy and potentially intractable $\hat U$, a
shorter $\hat U^{(j)}$ is prepared based on the current \gls{DIS}, its
parameters are optimized, and both $\hat U^{(j)}$ and optimal
$\boldsymbol\tau^{j}$ are used to transform (``dress'') the current
Hamiltonian $\hat H^{(j)}$ to define a new one,
\begin{equation}
  \label{eq:dressing}
  \hat H^{(j+1)} = \left(\hat U^{(j)}(\boldsymbol\tau^{j})\right)^\dagger \hat
  H^{(j)} \hat U^{(j)}(\boldsymbol\tau^{j}),
\end{equation}
for which the procedure of finding the \gls{DIS} is repeated. It was
numerically demonstrated that such a procedure eventually converged to
the exact ground-state energy even if a \emph{single} top-ranked
generator from the \gls{DIS} was used to build $\hat U^{(j)}$. Thus,
it is possible to use arbitrarily shallow quantum circuits at
expense of additional dressing steps carried out on a classical
computer and additional quantum measurements of of intermediate
$H^{(j)}$.

The major issue of the \gls{iQCC} method is the exponential
growth of intermediate Hamiltonians upon dressing, which severely
limits the problems size and/or the number of iterations possible in
the \gls{iQCC} procedure. Operator compression and energy
extrapolation was suggested in Ref.~\citenum{Ryabinkin:2020/jctc/1055} as mitigation techniques, but
both of them were only moderately efficient. Alternatively, a special
form of the \gls{QCC} ansatz~\eqref{eq:qcc_form}, which uses the
involutary combinations of anti-commuting generators, was recently
proposed~\cite{Lang:2020/arXiv/2002.05701}; it guarantees only
quadratic (with respect to the number of generators) increase of the
size of the Hamiltonian upon dressing.

In this work we explore a complementary approach. Leaving the
\gls{QCC} ansatz intact, we devise various \textit{a posteriori}
completeness corrections to the \gls{iQCC} energies, which can be
efficiently evaluated on a classical computer. In other words, we
introduce a classical post-processing technique that improves the
accuracy of the \gls{iQCC} energies or reduces the number of
\gls{iQCC} iterations for the same accuracy. Obviously, this saves
quantum resources. Moreover, capturing a bulk of contributions to the
exact energy classically allows us to consider the \gls{QCC}
form~\eqref{eq:qcc_form}, in which only a subset of all qubits is
considered explicitly, enabling the active-space treatment. This idea
has already been explored in the
literature~\cite{Takeshita:2020/prx/011004}. The distinctive feature
of our approach is that we simultaneously increase the accuracy
\emph{and} reduce the number of quantum measurements as compared to
the original \gls{iQCC} method.

The paper is organized as follows. First, we introduce a new form of
qubit Hamilonians that provides an algebraic point of view on the
\gls{DIS}. With the new form we discuss a closely related linear
variational ansatz for a qubit wavefunction, which becomes a starting
point for the derivation of a variety of variational and perturbative
corrections to the \gls{iQCC} method. The theoretical section ends
with the discussion of the active-space modification of \gls{iQCC}. In
the subsequent sections we assess the utility and performance of the
proposed modifications on three trial problems: 10-qubit \ce{N2}
molecule dissociation, the 24-qubit \ce{H2O} symmetric stretch, and
56-qubit singlet-triplet gap calculations for the
tris-(2-phenylpyridine)iridium(III), \ce{Ir(ppy)3} molecule.

\section{Theory}
\label{sec:theory}

\subsection{Algebraic definition of the \acrlong{DIS}}
\label{sec:algebr-defin-dis}

Any qubit Hamiltonian~\eqref{eq:qubitH} can always be written as
\begin{equation}
  \label{eq:ZX_part}
  \hat H = \sum_k {\tilde C}_k \hat Z_{k'} \hat X_k,
\end{equation}
where $Z_{k'} = z_{k'_1} \cdots z_{k'_m}$ and
$X_k = x_{k_1} \cdots x_{k_n}$ are Pauli words containing $\hat z$ or
$\hat x$ operators only. The representation~\eqref{eq:ZX_part} follows
from the fact that $\hat z$ and $\hat x$ operators together with the
imaginary unit $\I$ are the generators of the Pauli group, since
$\hat y = -\I \hat z \hat x$ and
$\hat x^2 = \hat y^2 = \hat z^2 = \hat e$, where $\hat e$ is the
identity operator. To obtain Eq.~\eqref{eq:ZX_part} from
Eq.~\eqref{eq:qubitH} one must replace all occurrences of $\hat y_j$
with $-\I \hat z_j \hat x_j$ in every $\hat P_k$ and collect
$\hat z_{k'_i}$ and $\hat x_{k_i}$ in factors $\hat Z_{k'}$ and
$\hat X_{k}$ respectively. The multi-indices $\{k'_1 \ldots k'_m\}$
and $\{k_1 \ldots k_n\}$ may overlap; the common indices correspond to
the $\hat y$ operators in these positions. The coefficients
${\tilde C}_k$ in Eq.~\eqref{eq:XZ_part} may differ from $C_k$ in an
additional phase factor $(\pm 1, \pm\I)$, depending on the number of
$\hat y$ factors in the corresponding $\hat P_k$.

Together with the ZX (``right'') expansion one may define an
alternative XZ (``left'') expansion as
\begin{equation}
  \label{eq:XZ_part}
  \hat H = \sum_k {\tilde C}'_k \hat X_{k'} \hat Z_k.
\end{equation}
The operator factors $\hat X_{k'}$ and $\hat Z_k$ are identical
between the ``right'' and the ``left'' variants, but the corresponding
coefficients may have different signs
${\tilde C}'_k = \pm {\tilde C}_k$. However, as long as all
$\hat P_k$-s contain the \emph{even} number of $\hat y$ factors, the
coefficients are identical. Such hermitian Hamiltonians with the even
number of $\hat y$ have \emph{real} matrix elements in a real basis
set. In fact, all electronic Hamiltonians in the absence of magnetic
fields and spin-orbital interaction are of this kind.

Expansion~\eqref{eq:ZX_part} can be regrouped as
\begin{equation}
  \label{eq:Ising_decomp}
  \hat H = {\hat I}_0(\mathbf{z}) + \sum_{k} {\hat I_k}(\mathbf{z}) X_k,
\end{equation}
where $I_k(\mathbf{z}) = I_k(\hat z_1, \hat z_2, \dots)$ are
generalized Ising Hamiltonians containing only the products of
$\hat z_j$ operators. We assume that $k$ is an integer whose binary
representation matches the string of Pauli elementary $\hat x$
operators such that 1 in the $i$-th position indicates the presence of
$\hat x_i$.

The motivation for treating $X$ and $Z$-dependent components
differently is explained below. Imagine that $\hat H$ acts on a
direct-product $n$-qubit wavefunction,
\begin{equation}
  \label{eq:product_wf}
  \ket{\Phi} = \ket{\pm 1}_1 \cdots \ket{\pm 1}_n,
\end{equation}
where $\ket{\pm 1}_j$ is the eigenstate of $\hat z_j$ operator with
eigenvalues $+1$ or $-1$. There are $2^n$ of such states which
correspond to distinct strings of $+1$ and $-1$. These
linearly-independent states were called the ``perfect mean-filed
states'' in Ref.~\citenum{Ryabinkin:2020/jctc/1055}. Every product
state is an eigenstate of an arbitrary Ising Hamiltonian, whereas
Pauli strings $\hat X_m$ map one mean-field state into another. That
is, the expectation value of an Ising Hamiltonian $I_k(\mathbf{z})$ on
an arbitrary mean-field state $\ket{0}$ is, in general, non-zero,
\begin{equation}
  \label{eq:omega_defs}
  \omega_k = |\braket{0|I_k(\mathbf{z})|0}| \ne 0,
\end{equation}
whereas
\begin{equation}
  \label{eq:X_av_pmf}
  \braket{0| \hat X_k |0} = 0, \ k>0
\end{equation}
Both properties are crucial for defining the \gls{DIS}. Consider a
single-generator ansatz~\eqref{eq:qcc_form},
\begin{equation}
  \label{eq:QCC_single}
  \hat U(\tau) = \exp(-\I \tau \hat P /2),
\end{equation}
and the expectation value of the canonically-transformed Hamiltonian
\begin{equation}
  \label{eq:QCC_single_energy}
  E(\tau) = \braket{0|\hat U^\dagger(\tau) H \hat U(\tau)|0}.
\end{equation}
The derivative of $E(\tau)$ with respect to $\tau$ is:
\begin{align}
  \label{eq:QCC_sngl_gen}
  \frac{\mathrm{d} E}{\mathrm{d}\tau} \Big|_{\tau = 0}
  & = -\frac{\I}{2}\Braket{0|[\hat H, \hat P]|0}
\end{align}
The \gls{DIS} comprises of all the operators satisfying
$\left|\frac{\mathrm{d} E}{\mathrm{d}\tau}\right| \ne 0$. Using the
expansion~\eqref{eq:qubitH} we can write:
\begin{equation}
  \label{eq:DIS_eq}
  \left|\frac{\mathrm{d} E}{\mathrm{d}\tau}\right| \le \sum_{k >0} \omega_k \left|\frac{1}{2\I}\braket{0|[\hat X_k, \hat P]|0}\right| =
  \sum_{k >0} \omega_k |\text{Im}\braket{0|\hat X_k \hat P|0}|
\end{equation}
Because terms of Eq.~\eqref{eq:DIS_eq} are algebraically independent,
at least one of them must be non-zero, which in view of
Eqs.~\eqref{eq:omega_defs} and \eqref{eq:X_av_pmf} implies
that~\footnote{%
  Strictly speaking, the condition must be written as
  $$
  \hat X_k \hat P = \hat I_{k'}(\mathbf{z})
  $$
  where $\hat I_{k'}(\mathbf{z})$ is a new generalized Ising
  Hamiltonian. It can be chosen arbitrarily, for example, to make the
  resulting generator commuting with global symmetry operators, such
  as the electron number $\hat N$ or the total spin-squared $\hat S^2$
  ones. Additionally, this operator can be made unimodular,
  \textit{i.e.} $\braket{0|\hat I_{k'}|0} = 1$.}
\begin{equation}
  \label{eq:P_cond}
  \hat X_k \hat P   = \hat Z_{k'},
\end{equation}
where $k'$ runs from 1 to $2^n$. To guarantee that the
\emph{imaginary} part of the l.h.s.\ of Eq.~\eqref{eq:P_cond} is
non-zero, $k'$ is additionally subjected to a condition that $k'$ and
$k$ must intersect in an \emph{odd} number of bits, which leaves
$2^{n-1}$ variants for $k'$. Eq.~\eqref{eq:P_cond} can be easily
solved by multiplying both sides by $\hat X_k$:
\begin{equation}
  \label{eq:P_DIS}
  \hat P_{kk'} = \hat X_k \hat Z_{k'}.
\end{equation}
Such defined $\hat P_{kk'}$ will contain the odd number of $\hat y$
and satisfy
\begin{equation}
  \label{eq:P-grad}
  \left | \frac{\mathrm{d} E[\hat P_{kk'}]}{\mathrm{d}\tau} \right|  = \omega_k
  |\braket{0|Z_{k'}|0}| = \omega_k.
\end{equation}

Thus, the \gls{DIS} is a set of solutions of Eq.~\eqref{eq:P_cond}
with indices $k$ coming from the expansion~\eqref{eq:Ising_decomp} and
$k'$ subjected the condition above. Different $k$ characterize
different \emph{groups} of operators in the \gls{DIS} with different
values of gradients $\omega_k$, whereas different $k'$ label operators
with the \emph{identical} gradients; in other words, the \gls{DIS} can
be described as a union of groups of operators. In
Ref.~\citenum{Ryabinkin:2020/jctc/1055} the binary representation of
each $k$ was called ``the flip set'' with respect to an ideal
mean-field reference $\ket{0}$. It is now clear that flip sets are
binary representations of $X_k$ in the
expansion~\eqref{eq:Ising_decomp}, and thus, reference-independent.

We emphasize that \gls{DIS} is properly defined only for the
mean-field references although the energy gradients can be evaluated
by Eq.~\eqref{eq:QCC_sngl_gen} for any reference. However, for a
general non-mean-field state $\ket{0}$ the
condition~\eqref{eq:X_av_pmf} no longer holds, and operators outside
the \gls{DIS} acquire non-zero gradient values. We also note that the
conditions~\eqref{eq:omega_defs} and \eqref{eq:X_av_pmf} are analogs
of the Wick's theorem for the ``normal ordered'' qubit
expression~\eqref{eq:Ising_decomp}.

\subsection{Linear variational ansatz for a qubit wavefunction}
\label{sec:line-vari-ansatz}

There are exactly $2^n$ distinct Pauli strings $\hat X_k$ for $n$
qubits. Since from $\hat X_k \ket{0} = \hat X_m \ket{0}$ follows that
$\hat X_k = \hat X_m$ for an arbitrary but fixed mean-field state
$\ket{0}$, \emph{any} mean-filed state can be represented as
$\hat X_k \ket{0}$ for some $k$. The mean-field states for $n$ qubits
form a complete basis in the Hilbert space, so any $n$-qubit wave
function can be expressed as
\begin{equation}
  \label{eq:linear_variational_ansatz}
  \ket{\Psi} = \left(1 + \sum_{k>0} d_k \hat X_k\right) \ket{0}
\end{equation}
(the intermediate normalization $\braket{0|\Psi} = 1$ is assumed). The
linear variational ansatz~\eqref{eq:linear_variational_ansatz} is
naturally associated with the matrix eigenvalue problem
\begin{equation}
  \label{eq:CI_problem}
  \mathbf{H} \mathbf{d} = E \mathbf{d},
\end{equation}
with the matrix
$\mathbf{H} = \{\braket{0|\hat X_i \hat H \hat X_j|0}\}_{i,j =
  0}^{M}$, where $M \le 2^n$; $\mathbf{d} = \{d_k\}_{k = 0}^M$ is an
eigenvector.

Equation~\eqref{eq:CI_problem} can be solved on a \emph{classical} computer
to get corrected ground-state energy. Moreover, if one includes all
$2^n$ Pauli $X$ strings, this will give the exact ground-state energy
and wavefunction. This procedure, however, is equivalent to \gls{FCI}
and, hence, intractable. A more convenient way is to include all
directly coupled to $\ket{0}$ mean-field states. Consider the 0-th row
of the Hamiltonian matrix $\mathbf{H}$:
\begin{equation}
  \label{eq:H0j}
  H_{0j} = \braket{0|\hat H \hat X_j|0}.
\end{equation}
If these matrix elements are required to be non-zero, than the set of
the corresponding $\hat X_j$ coincides with the set of $\hat X_k$
entering the expansion~\eqref{eq:Ising_decomp}; in other words,
matches the group structure of the \gls{DIS}. In fact, the whole row
$H_{0j}$, $j > 0$ is evaluated as a part of the ranking procedure at
each \gls{iQCC} iteration~\cite{Ryabinkin:2020/jctc/1055}. To complete
the construction of the Hamiltonian matrix in this case one needs to
compute remaining matrix elements
$\braket{0|\hat X_i \hat H \hat X_j|0}$ for $i,j > 0$ and
$i, j \in \text{DIS}$. Since the size of the \gls{DIS} is at worst
linear in the size of the qubit Hamiltonian, this procedure is
computationally feasible.

\subsection{\acrfull{ENPT} for a qubit wavefunction}
\label{sec:epst-nesb-pert}

The variational improvement of the \gls{iQCC} energies by
Eq.~\eqref{eq:CI_problem} may still be too computationally demanding
since the size of the \gls{iQCC} dressed Hamiltonians rapidly
increases. Moreover, when \gls{iQCC} energies approach the exact,
corrections become smaller, which naturally calls for
perturbation-theory consideration. The simplest solution is based on
the \gls{ENPT}, in which all the off-diagonal couplings~\eqref{eq:H0j}
are treated as perturbations due to the presence of mean-field states
$\ket{j} = \hat X_j \ket{0}$ with energies
\begin{equation}
  \label{eq:E_k}
  E_j = \braket{j| \hat H |j} = \braket{0|\hat X_j \hat H \hat X_j|0}
  = \braket{j| \hat I_0(\mathbf{z}) |j}.
\end{equation}
The second-order \gls{ENPT} energy correction formula
is~\cite{Epstein:1926/pr/695, Nesbet:1955/prsl/312}:
\begin{equation}
  \label{eq:ENPT2_energy}
  \Delta E_\text{ENPT}^{(2)} = -\sum_j \frac{|H_{0j}|^2}{E_j - E_0} =
  -\sum_j \frac{\omega_j^2}{D_j},
\end{equation}
where $\omega_j$ are the absolute values of the \gls{QCC} energy
derivatives, Eq.~\eqref{eq:P-grad}, and
\begin{equation}
  \label{eq:PT_denominators}
  D_j = E_j - E_0
\end{equation}
are energy denominators. Energy derivatives $\omega_j$ are readily available at the end
of each \gls{iQCC} iterations and the only extra quantities that need
to be computed are $D_j$. The computational overhead is strictly
linear in the size of the \gls{DIS}, which has to be contrasted to the
qubic scaling of computational efforts associated with the matrix
diagonalization problem Eq.~\eqref{eq:CI_problem}.

\subsection{Diagonal unitary (infinite-order) modification of the
  \gls{ENPT}2 correction}
\label{sec:diag-infin-order}

The second-order \gls{ENPT} correction Eq.~\eqref{eq:ENPT2_energy}
diverges when $D_j \to 0$. The problem can be solved by modifying the
denominators, and a variety of strategies were suggested (see
Refs.~\citenum{Li:2019/arpc/245} and \citenum{Park:2019/jctc/4088} and
references therein). Here we derive our own variant which is more
aligned with the \gls{QCC} approach. Consider a qubit Hamiltonian with
only one $X$ term, $\hat I_k(\mathbf{z}) \hat X_k$, in
Eq.~\eqref{eq:Ising_decomp}. This is the case, for example, of the
\ce{H2} molecule in the minimal basis. It is easy to show that a
single-generator \gls{QCC} ansatz with $\hat P_k = \hat X_k \hat Z_j$,
where $j$ is an arbitrary number from 1 to $2^n$ having the odd-number
bit overlap with $k$, finds the exact ground state. Indeed, the
variational expression for the energy is:
\begin{align}
  \label{eq:E_QCC_single}
  E(\tau)  = {} & \braket{0|\E^{\I \tau \hat P_k/2} \hat H \E^{-\I \tau
                  \hat P_k/2}|0 } \\ \nonumber
  = {} & E_0 + \Braket{0|\left(\frac{1}{2\I}\right) [\hat H, \hat P_k]|0} \sin\tau \\ \nonumber
  {}  & {} + \Braket{0|\frac{1}{2}\left(\hat P_k \hat H \hat P_k - H\right)|0} \left(1 - \cos\tau\right),
\end{align}
where $E_0 = \braket{0| \hat H |0}$. Identifying $\omega_k = \braket{0|I_k(\mathbf{z})|0}$ and
$D_k = \braket{0|\hat P_k \hat H \hat P_k - H|0} = E_k - E_0$ we can write
\begin{equation}
  \label{eq:E-trig}
  E(\tau) = E_0 + \omega_k \sin{\tau} + \frac{D_k}{2}\,
  \left(1-\cos{\tau}\right).
\end{equation}
Combining all trigonometric functions together we find that 
\begin{equation}
  \label{eq:E-trig_final}
  E(\tau) = E_0 + \frac{D_k}{2} -
  \sqrt{\left(\frac{D_k}{2}\right)^2+\omega_k^2}\,\cos(\tau
  + \phi_k),
\end{equation}
where $\phi_k = \arcsin\frac{2\omega_k}{\sqrt{D_k^2 + 4\omega_k^2}}$.
The minimum of the Eq.~\eqref{eq:E-trig_final} corresponds to
$\cos(t + \phi_k) = 1$, and the energy correction is:
\begin{equation}
  \label{eq:DUC_basic}
  E_\text{gs} - E_0 = \frac{D_k}{2} - \sqrt{\left(\frac{D_k}{2}\right)^2 + \omega_k^2} .
\end{equation}

It is clear that $\omega_k^2$ and $D_k$ are the numerator and the
denominator of one particular term in
Eq.~\eqref{eq:ENPT2_energy}. Thus, assuming independent (fully
uncorrelated) contribution of every generator $\hat P_k$ we can write
\begin{equation}
  \label{eq:DUC}
  \Delta E_\text{DUC} = \sum_k \left(\frac{D_k}{2} -
    \sqrt{\left(\frac{D_k}{2}\right)^2 + \omega_k^2} \right),
\end{equation}
where DUC stands for the ``diagonal unitary correction''. In the case
of large denominators $E_\text{DUC}$ reduces to the \gls{ENPT}2
expression~\eqref{eq:ENPT2_energy} but remains finite when any
$D_k \to 0$. A similar modification of the second-order
M\/{o}ller--Plesset perturbation correction has been introduced in
Ref.~\citenum{Assfeld:1995/cpl/438}.

\subsection{Combined variational-perturbative correction}
\label{sec:comb-vari-pert}

The variational and perturbative corrections to the \gls{QCC} method
introduced above have their own strengths and weaknesses. The
variational approach is considerably more demanding computationally
but is superior if the ground state becomes quasi-degenerate with a
lower excited state. The situation is not uncommon in molecular
systems and is known as conical intersections~\cite{Migani:2004/271,
  Yarkony:1996/rmp/985}. On the other hand, the perturbation
correction is especially convenient as many quantities are already
computed during the \gls{iQCC} iteration and no diagonalization is
required. Below we suggest a solution that combines the strengths of
both approaches, namely, the ability to cope with low-lying
quasidegeneracies and efficient (perturbative) treatment of high-lying
states. We commence with the partitioning of the full matrix
problem~\eqref{eq:CI_problem} into smaller sub-problems in the spirit
of the L\"owding partitioning~\cite{Lowdin:1963/jms/12} as
\begin{align}
  \label{eq:eig_part_eq1}
  \mathbf{h} \mathbf{p} + \mathbf{b}^\dagger \mathbf{q} & = E \mathbf{p} \\
  \label{eq:eig_part_eq2}
  \mathbf{b} \mathbf{p} + \mathbf{C} \mathbf{q} & = E \mathbf{q},
\end{align}
where $\mathbf{h}$ is a $m \times m$ submatrix of $\mathbf{H}$,
$\mathbf{b}^\dagger$ is the remaining upper right part, and
$\mathbf{C}$ is a matrix with the mean-field energies~\eqref{eq:E_k}
on diagonals. $1 \le m \le N+1$, where $N$ is the number of groups in
the \gls{DIS}. The first state is the ground-state reference
$\ket{0}$. $\mathbf{p}$ and $\mathbf{q}$ are first $m$ and remaining
$N+1-m$ components of the full eigenvector $\mathbf{d}$ [see
Eq.~\eqref{eq:CI_problem}], respectively. If $E$ is an eigenstate, we
can solve Eq.~\eqref{eq:eig_part_eq2} for $\mathbf{q}$
and plug it into Eq.~\eqref{eq:eig_part_eq1} to obtain
\begin{align}
  \label{eq:eff_eigproblem}
  \mathbf{h}_\text{eff}(E) \mathbf{p} & = E \mathbf{p},
\end{align}
where
\begin{equation}
  \label{eq:Heff}
  \mathbf{h}_\text{eff}(E) = \mathbf{h} + \boldsymbol\Sigma(E).
\end{equation}
is the matrix of the effective energy-dependent Hamiltonian, and
\begin{equation}
  \label{eq:self-energy}
  \boldsymbol\Sigma(E) = -\mathbf{b}^\dagger (\mathbf{C} - E)^{-1}
  \mathbf{b}
\end{equation}
is the self-energy. If $m = N+1$ than the self-energy vanishes and the
problem~\eqref{eq:eff_eigproblem} reduces to the original matrix
formulation~\eqref{eq:CI_problem}. In the opposite limit, $m=1$,
Eq.~\eqref{eq:eff_eigproblem} is a non-linear equation, which,
nevertheless, can be solved for the exact ground-state energy provided
that the self-energy is exact. However, as follows from
Eq.~\eqref{eq:self-energy}, this amounts to full inversion of the
$(\mathbf{C}-E)$ matrix, which scales cubicly in the matrix dimension, similar to the diagonalization. 
The inversion is trivial though, if only the
diagonal matrix elements of $\mathbf{C}$ are retained; the self-energy
can be computed in this case via simple multiplication of
$\mathbf{b}^\dagger$ and $(\mathbf{C} - E)^{-1}\mathbf{b}$. Working
out this product explicitly, we find:
\begin{equation}
  \label{eq:SE_1}
  \boldsymbol\Sigma(E) = -\sum_j \frac{H^{*}_{0j}H_{j0}}{E_j - E} =
  -\sum_j \frac{\omega_j^2}{E_j - E}.
\end{equation}
The only difference between this formula and
Eq.~\eqref{eq:ENPT2_energy} is the use of the exact (corrected) energy
$E$ in place of $E_0$. Equation~\eqref{eq:SE_1} turns out to be the
(second-order) Brillouin-Wigner perturbation
theory~\cite{Lennard-Jones:1930/prsca/598, Brillouin:1932/jpr/373,
  Wigner:1937/collection/131} correction. Since $E$ is not initially
known, iterations are necessary to compute the final value. However,
contrary to its \gls{ENPT} counterpart, the Brillouin-Wigner
perturbation theory is not susceptible to divergence due to vanishing
denominators. In this regard it is a competitor to the diagonal
unitary correction introduced in Sec.~\ref{sec:diag-infin-order}.

The intermediate case $m > 1$ with the diagonal approximation for
$\mathbf{C}$ can be thought of as a multiconfigurational perturbation
theory, which bears some similarity with known
variants~\cite{Nakano:1993/cpl/372}. The coupling among $m$ states is
computed exactly, as well as their coupling to the remaining $N+1-m$
``external'' states; only the coupling between external states is
neglected. The number of states $m$ can be chosen based on efficiency or accuracy
considerations. For example, $m$ can be set statically to mitigate the
problem of enlarging intermediate \gls{iQCC} Hamiltonians or be
adjusted dynamically, to include the states that still
have appreciable couplings with the reference state.

\subsection{Perturbative generators' ranking for \gls{iQCC}}
\label{sec:pert-rank-gener}

The \gls{iQCC} method ranks generators $\hat P_{kk'}$ according to
their absolute gradients $\omega_k$, see Eqs.~\eqref{eq:omega_defs}
and \eqref{eq:P-grad}. Operators associated with highest gradients are
selected first for including into the \gls{QCC} ansatz at the next
iteration. If any of the corrections introduced in
Sec.~\ref{sec:line-vari-ansatz}--~\ref{sec:comb-vari-pert} are going
to be computed, alternative rankings become available. Namely, one can
consider the first-order \gls{ENPT} contribution to the wavefunction,
\begin{equation}
  \label{eq:PT1_coeff}
  |d_\text{j, ENPT}^{(1)}| = \frac{2\omega_j}{|D_j|}
\end{equation}
or the second-order \gls{ENPT} absolute energy increments,
\begin{equation}
  \label{eq:PT2_energy_cont}
  |\Delta E_\text{j, ENPT}^{(2)}| = \frac{\omega_j^2}{|D_j|}.
\end{equation}
The first formula can be advantageous in the case of near-degeneracy,
in which small energy changes are accompanied by large amplitude
variations. Equation~\eqref{eq:PT1_coeff} directly estimate the amplitude
by dividing a gradient value by an energy gap. Unfortunately, since
the transition between weak and strong correlation regimes can be
smooth to finally gauge which ranking is preferable the numerical
testing is required. We perform such comparison in the subsequent
sections.

\subsection{Active-space treatment in \gls{iQCC}}
\label{sec:active-space-treatm}


\gls{iQCC} is a variational method, which is equally capable of
handling cases of weak and strong correlation. However, the canonical
transformation step~\eqref{eq:dressing} invariably leads to expansion
of intermediate Hamiltonians. It is important, therefore, to carefully
control which generators are included into the \gls{QCC}
ansatz~\eqref{eq:qcc_form}. One way is to employ alternative ranking
schemes, like those suggested in Sec.~\ref{sec:pert-rank-gener}. A
more direct approach is to split qubit indices into active and
inactive sets. The active indices are assigned to a subsystem with
strong mixing (entanglement), while the remaining ones are spectators
which can be treated approximately. The most straightforward
implementation of the idea can be made using the \acrlong{JW}
fermion-to-qubit mapping, as in this case there is one-to-one
correspondence between spin-orbitals and qubit indices, so that the
traditional \gls{CASSCF} treatment~\cite{Helgaker:2000} will provide
the guidance how to select the active qubit indices. Generators whose
qubit indices are fully within the active set can be called
``internal''; only such generators are included into the \gls{QCC}
ansatz. The other, ``external'' generators can be handled by
perturbative/variational corrections introduced above. The
construction can be made even more versatile by including
``semi-internal'' generators with a predefined number (1, 2, or more)
of inactive indices into the \gls{QCC} form. These semi-internal
operators account for partial relaxation of the environmental
(external) \acrlongpl{DOF}. However, generators whose indices are
fully external will never be treated exactly making the whole approach
approximate.

\section{Notes on implementation}
\label{sec:implementation}

All the corrections developed above can be easily integrated into the
\gls{iQCC} workflow. After the \gls{iQCC} procedure has completed the
optimization of amplitudes in a chosen \gls{QCC} form using the energy
values sampled by a quantum computer, the Hamiltonian is dressed by
Eq.~\eqref{eq:dressing} on a classical computer using the optimized
amplitudes. Subsequently, for the new Hamiltonian, representatives and
the corresponding gradients for each group from its \gls{DIS} are
generated by Eqs.~\eqref{eq:P_DIS} and \eqref{eq:P-grad},
respectively. The new step is calculating the excited mean-field
energies by Eq.~\eqref{eq:E_k}; this procedure has computational
complexity comparable to the previous \gls{DIS} computations. Two out
of three perturbative corrections can be computed immediately by
Eqs.~\eqref{eq:ENPT2_energy} and \eqref{eq:DUC} with negligible cost.
The Brillouin-Wigner correction [Eq.~\eqref{eq:SE_1}] requires an
additional self-consistent procedure, whose cost is still small
compared to the gradient computations. The variational-perturbative
correction with the effective Hamiltonian of the dimension $m >1$
($m \le N+1$, where $N$ is the number of groups in the \gls{DIS})
requires calculation of $(m-1)(m-2)/2 + (m-1)(N+1-m)$ extra matrix
elements of the matrix $\mathbf{H}$ [see Eq.~\eqref{eq:H0j}] plus
multiple diagonalizations of $\mathbf{h}_\text{eff}$ [see
Eq.~\eqref{eq:Heff}] to reach self-consistency. This procedure can be
time- and memory-consuming, especially for large $m$ and late
\gls{iQCC} iterations, so that it is advised only if severe state
degeneracy (\textit{i.e.} multiple bond breaking) is anticipated.

Finally, the specified number of top-ranked generators are selected
for the next iteration. Ranking is based on the absolute values of
gradients (the original \gls{iQCC} prescription) or using the
measure~\eqref{eq:PT1_coeff}. At this step the active-space treatment
can be engaged by requiring that generators must belong to the active
space. If the maximum gradient associated with chosen generators is
below the convergence threshold, the \gls{iQCC} procedure is stopped,
otherwise a the new iteration is started. Convergence may also be
declared if a variation of the corrected \gls{iQCC} energies between
two successive iterations drops below a specified threshold.

\section{Results and discussion}
\label{sec:results-discussion}

All the numeric results reported here are obtained
classically. The quantum part of the \gls{iQCC} algorithm, namely, the
optimization of amplitudes in the \gls{QCC} ansatz at each \gls{iQCC}
iteration is performed on a classical computer. We set aside,
therefore, any problems with noisy optimization of amplitudes;
however, the whole \gls{iQCC} scheme does not rely on the precise
optimization of them. Even if only slightly improved (lower) energies
are obtained, the procedure can move forward albeit at reduced
efficiency. Dressing, building the \gls{DIS}, (re)computing of the
\gls{iQCC} energies and corrections are invariably made on a classical
computer. Thus, the quantum device is to be used as a ``quantum
accelerator'', much like how the graphical processing units (GPU) are
currently used to speed up certain steps of electronic structure
calculations~\cite{Fales:2020/jctc/1586}.

First we investigate one of the fundamental properties of any
electonic structure method---the size consistency. The size
consistency implies the correct (linear) scaling of energies when
multiple \emph{non-interacting} subsystems are considered. The
\gls{iQCC} method itself is size-consistent. It is known, however,
that neither the linear variational method, nor the Epstein--Nesbet or
the Brillouin--Wigner perturbation theories are
size-consistent~\cite{Malrieu:1979/tca}. We expect, therefore, the
lack of size consistency for our corrections too. This anticipated
flaw, however, is not severe since the corrections to the \gls{iQCC}
energies (and hence the size-consistency error) can be made
arbitrarily small at expense of additional iterations.

On the 10-qubit example of \ce{N2} dissociation we assess the ability
of the corrections to reduce the number of \gls{iQCC} iterations for
the same target accuracy in both weakly and strongly correlated
regimes. The 24-qubit symmetric water molecule stretching problem
illustrates the active-space treatment and, finally, the large
(56-qubit) problem of a single-triplet gap in \ce{Ir(ppy)3}
demonstrates the scalability of the corrected \gls{iQCC} method.

\subsection{Size-consistency test: A non-interacting \ce{(H2)2}}
\label{sec:size-cons-pert}

We consider the non-interacting hydrogen molecule dimer \ce{(H2)2} in
a planar rectangular geometry, in which the individual \ce{H2}
moieties with $\ce{R(H-H)} = \SI{0.75}{\angstrom}$ are
\SI{100}{\angstrom} far apart. The electronic Hamiltonians in the
minimal STO-6G basis set are mapped by the \gls{JW} transformation to
4- and 8-qubit operators for \ce{H2} and \ce{(H2)2}, respectively. As
was noted above, the \gls{JW} mapping allows for exceptionally simple
interpretation of a qubit mean-field wavefunction: each occupied
spin-orbital is mapped to a $\ket{-1}$ state of the corresponding
qubit.

Pauli words and coefficients in Eq.~\eqref{eq:qubitH} depend also on a
\gls{MO} set. For a \ce{H2} the natural choice is the set of the
canonical Hartree--Fock orbitals; in this case the resulting 4-qubit
Hamiltonian has only two terms in Eq.~\eqref{eq:Ising_decomp}:
\begin{equation}
  \label{eq:qubitHam_H2}
  \hat H(\ce{H2}) = \hat I_0(\mathbf{z}) + I_{15}(\mathbf{z})\, {\hat x}_3
  {\hat x}_2 {\hat x}_1 {\hat x}_0,
\end{equation}
that is, the \gls{DIS} has only one group. The qubit operator
${\hat x}_3 {\hat x}_2 {\hat x}_1 {\hat x}_0$ acting on the product
state
$\ket{\mathbf{0}} = \ket{-1}_0 \ket{-1}_1 \ket{+1}_2
\ket{+1}_3$~\footnote{The occupied Hartree--Fock spin-orbitals
  $\ket{1\alpha}$ and $\ket{1\beta}$ are mapped to single-qubit states
  $\ket{-1}_0$ and $\ket{-1}_1$, respectively.} maps it to the state
$\ket{+1}_0 \ket{+1}_1 \ket{-1}_2 \ket{-1}_3$; in other words, it is a
double-excitation operator. Mixing the doubly-excited configuration
with $\ket{\mathbf{0}}$ defines the \gls{FCI} problem for a singlet
state in this minimal basis set. The \gls{iQCC} method converges at
the second iteration after a single dressing of the initial
Hamiltonian~\eqref{eq:qubitHam_H2} with
$\exp(-\I \tau {\hat x}_3 {\hat x}_2 {\hat x}_1 {\hat y}_0/2)$. Since
the \gls{iQCC} procedure may take several iterations to converge for
the dimer, to make a fair comparison, we report the \gls{iQCC}
energies and their corrected values [by Eqs.~\eqref{eq:ENPT2_energy},
\eqref{eq:DUC}, and \eqref{eq:SE_1}] after the first iteration. Thus,
the corrections are applied directly to the Hartree--Fock state.

There are several choices of \gls{MO} sets for the dimer. The natural
one is the canonical (fully delocalized) Hartree--Fock \glspl{MO}.
Alternatively, one can consider a set of localized orbitals which are
sums of Hartree--Fock orbitals of \ce{H2} fragments. Since there is no
interaction between very distant fragments, the initial energy for
both sets,
$E_\text{iQCC}^{(1)} = \braket{\mathbf{0}|\hat H|\mathbf{0}}$ is
identical; see Table~\ref{tab:H2_2}.

\begin{table*}
  \centering
  \caption{Ground-state electronic energies (in \si{\hartree}) for the
    non-interacting hydrogen molecule dimer, \ce{(H2)2}, and their
    deviations from the corresponding doubled monomer energies in
    different approximation using various \gls{MO} sets. The iQCC
    energy is for the first iteration, where it equals to the average
    value of $\hat H$ on the qubit product state,
    $\braket{\mathbf{0}|\hat H|\mathbf{0}}$. The second-order
    Epstein--Nesbet (EN2), the diagonal unitary (DUC), and the
    Brillouin--Wigner (BW) corrections are computed by
    Eqs.~\eqref{eq:ENPT2_energy}, \eqref{eq:DUC}, and \eqref{eq:SE_1}
    respectively.}
  \begin{minipage}{1.0\linewidth}
    \sisetup{table-format = 3.6, %
      round-mode = places, %
      round-precision = 6} %
    \begin{tabularx}{1.0\linewidth}{@{}XSSSScccc@{}}
      \toprule
      Molecular orbital set                 &
                                              \multicolumn{4}{c}{Total energy}   & \multicolumn{4}{c}{Deviation, $E[\ce{(H2)2}] -  2E[\ce{H2}]$} \\
      \cmidrule{2-5}                                                 \cmidrule{6-9}
                                            & {iQCC}        & {+EN2}         & {+DUC}                        & {+BW}         & {iQCC}                                   & {+EN2}                                 & {+DUC}                              & {+BW} \\[0.75ex]
      Canonical Hartree--Fock               & -2.2494614989 &  -2.2824521803 & -2.2823847513                 & -2.2819276242 &\quad ${} < \num[round-precision=1]{e-9}$ &\quad \num[round-precision=1]{9.60E-03} & \num[round-precision=1]{9.10E-03}   & \num[round-precision=1]{9.56E-03} \\
      Fragment Hartree--Fock                & -2.2494614989 &  -2.2920505739 & -2.2914833452\footnotemark[1] & -2.2909449976 &\quad ${} < \num[round-precision=1]{e-9}$ & ${} < \num[round-precision=1]{e-9}$    & ${} < \num[round-precision=1]{e-9}$ & \num[round-precision=1]{5.38E-04} \\[1ex]
                                            & \multicolumn{8}{c}{Monomer} \\[0.75ex]

      Canonical Hartree--Fock               & -1.1247307495 & -1.1460252869  &-1.1457416726\footnotemark[1]  & -1.1457416726\footnotemark[1] & & & & \\
      \bottomrule
    \end{tabularx}
    \footnotetext[1]{The exact \gls{FCI} value.}
  \end{minipage}
  \label{tab:H2_2}
\end{table*}
From Table~\ref{tab:H2_2} it is clear that all corrections are
\emph{not} size-consistent, which is totally expected. The \gls{ENPT}2
and DUC forms [Eqs.~\eqref{eq:ENPT2_energy} and Eq.~\eqref{eq:DUC}],
but not the Brillouin--Wigner [Eq.~\eqref{eq:SE_1}] one become
size-consistent in the basis of fragment-local orbitals. On the other
hand, the diagonal unitary correction and the Brillouin--Wigner
formulas are exact for the two-level problem. Since the DUC and BW
corrections are also not prone to divergence due to small
denominators, they should be preferred over the \gls{ENPT}2 one.

\subsection{\ce{N2} dissociation}
\label{sec:cen2-dissociation}

The main goal of all corrections is boosting the computational
efficiency of the \gls{iQCC} method without sacrificing accuracy.
Since both pristine and corrected \gls{iQCC} energies ultimately
converge to the exact answer, an important characteristics is the
number of iterations that could be saved by applying corrections when
a certain accuracy threshold is targeted, for example,
\SI{0.1}{\milli\hartree}. To cover cases of weak and strong
correlation we consider the \ce{N2} dissociation curve using the
minimal 6-electron/6-orbital complete active space, CAS(6,6), which
allows for the correct dissociation of triply-bonded systems. In the
beginning, the canonical Hartree--Fock \glspl{MO} expanded in the
correlation-consistent double-zeta Dunning basis set
(cc-pVDZ)~\cite{Dunning:1989/jcp/1007} were generated for each
internuclear distance, and one- and two-electron integrals were
transformed to this \gls{MO} basis by the modified \textsc{gamess}
program~\cite{gamessus, gamessus-2}. The resulting second-quantized
electronic Hamiltonian was converted to a qubit form using the parity
transformation~\cite{Nielsen:2005/scholar_text}. The advantage of the
parity transformation used here is the presence of two stationary
qubits that can be removed; as a result, a 10-qubit effective
Hamiltonian containing the lowest singlet state of \ce{N2} at all
geometries can be constructed. The size of the problem, therefore, is
small enough to carry out \gls{iQCC} calculations with arbitrary
accuracy.

The \gls{iQCC} calculations were organized as follows. At every
iteration 4 generators with the largest EN1 contributions
[Eq.~\eqref{eq:PT1_coeff}] were selected for the use in the \gls{QCC}
ansatz~\eqref{eq:qcc_form} at the next one. Iterations were continued
until the absolute maximum gradient of the generators from the
\gls{DIS} [Eq.~\eqref{eq:P-grad}] decreased to 0.001, which translates
into $10^{-5}{-}10^{-6}$ \si{hartree} of accuracy in total energies.
Alternatively, the generators from the \gls{DIS} were ranked according
to their absolute gradient values, as in the original \gls{iQCC}
method. Note that the convergence criterion was chosen to be
independent on the ranking formula because different ranking
quantities have different physical dimensionality.

\begin{figure}
  \centering %
  \includegraphics[width=1.0\linewidth]{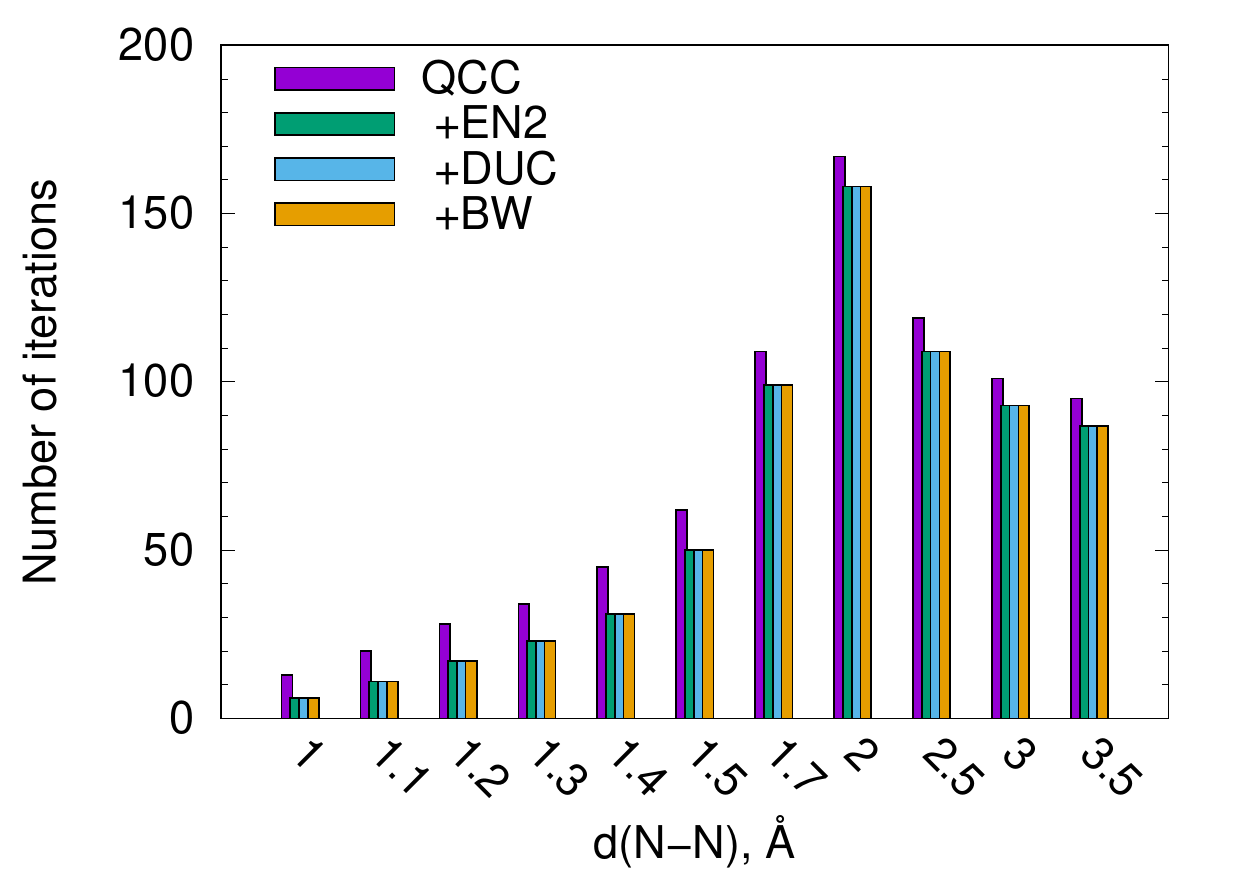}
  \caption{The number of iterations for the energy discrepancy
    $(E - E_\text{FCI})$ to fall below \SI{0.1}{\milli\hartree} as a
    function of $\ce{N-N}$ distance. Bare \gls{iQCC} energies and
    their corrected by Eqs.~\eqref{eq:ENPT2_energy}, \eqref{eq:DUC},
    and \eqref{eq:SE_1} counterparts labelled as ``EN2'', ``DUC'', and
    ``BW'', respectively.}
  \label{fig:N2_hist}
\end{figure}
Figure~\ref{fig:N2_hist} shows the number of iterations that are
necessary to reduce the energy discrepancy $|E - E_\text{FCI}|$ below
\SI{0.1}{\milli\hartree}. It is remarkable that at the chosen level of
accuracy all corrections require the same number of iterations. Near
the equilibrium geometry, $\ce{d(N-N)} \approx \SI{1.1}{\angstrom}$,
the corrections reduce the number of iterations by approximately a
factor of 2 (\textit{e.g.} at $\ce{d(N-N)} = \SI{1.0}{\angstrom}$ from
13 to 6). As the molecule is stretched, the efficiency of corrections
deteriorates. This result is not unexpected since the perturbative
treatment is inefficient in the strongly-correlated regime.
Nevertheless, the practical utility of the corrections is still
warranted as many experimental quantities, such as vibrational
frequencies, \textit{etc.} refer to near-equilibrium configurations.

\begin{figure}
  \centering
  \includegraphics[width=1.0\linewidth]{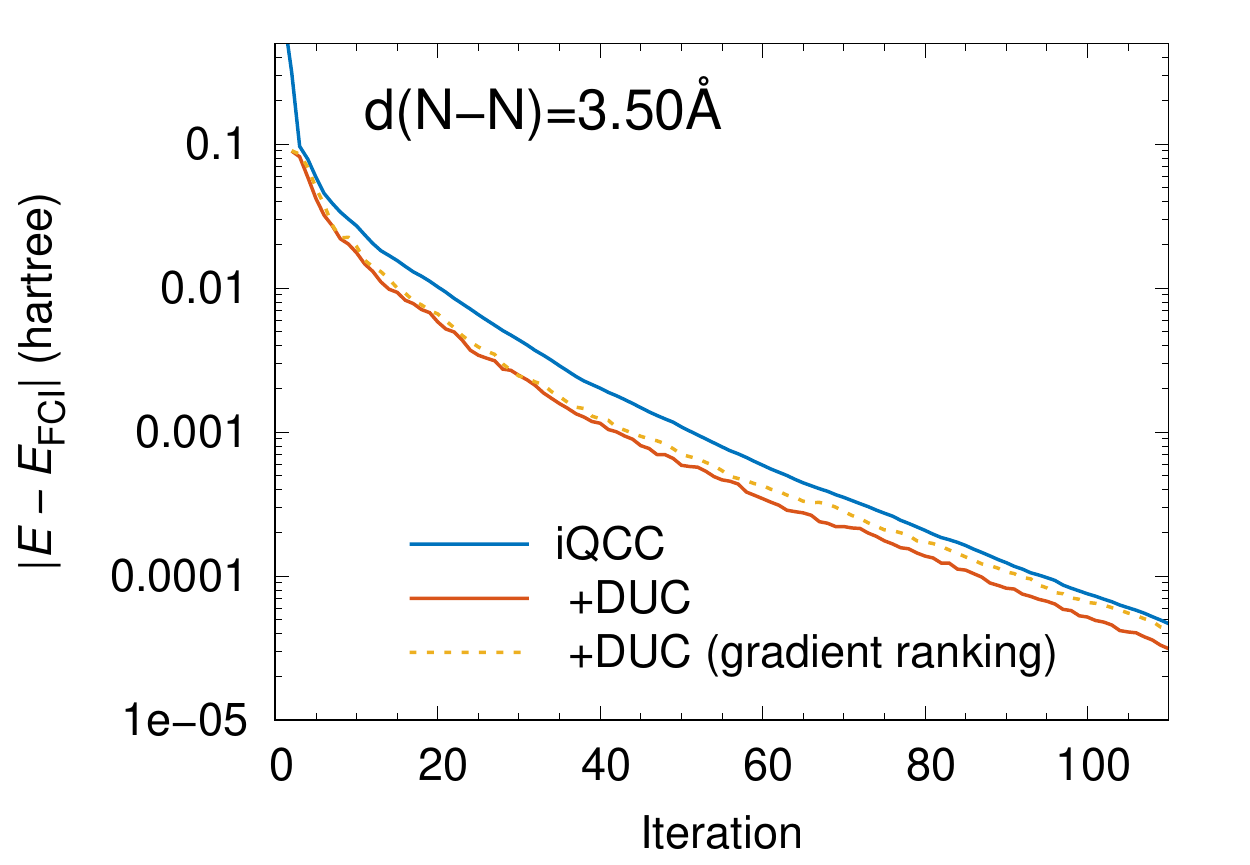}
  \caption{The absolute energy discrepancy $|E - E_\text{FCI}|$ as a
    function of iteration number for the bare \gls{iQCC} energies and
    their corrected by Eq.~\eqref{eq:DUC} values. Solid lines
    correspond to the EN1-based generator ranking in the \gls{iQCC}
    procedure, Eq.~\eqref{eq:PT1_coeff}, while the short dashed line
    uses the original, gradient-based one.}
  \label{fig:N2_en1_vs_grad}
\end{figure}
We also assessed the EN-based generator ranking procedure
[Eq.~\eqref{eq:PT1_coeff}]; see Fig.~\ref{fig:N2_en1_vs_grad}. As
anticipated, the EN-based ranking is better than the original,
gradient-based one, albeit marginally. Thus, we recommend the use of
the former for highly stretched molecules or in other cases of strong
correlation. Ranking by [Eq.~\eqref{eq:PT1_coeff}] has a negligible
computational cost as the weights are free by-products of the energy
corrections.

\subsection{Active-space treatment: The symmetric stretch of \ce{H2O}}
\label{sec:active-space-modif}

A symmetric water molecule stretch is another archetypal type of a
strong correlation problem. We consider a water molecule with fixed
$\angle\ce{HOH} = \SI{107.6}{\degree}$. When the 6-31G basis set is
used and 1s core orbital of an O atom is frozen, the electronic
Hamiltonian can be mapped to a 24-qubit operator. The size of the
Hilbert space is $2^{24} \approx \num{1.7e7}$, which makes the
straightforward \gls{iQCC} calculations difficult. We partition 24
qubits into two sets: the 8-qubit active one and the inactive set
containing the remaining qubits. We employed the \gls{JW}
fermion-to-qubit transformation, in which it is easy to link the
active qubit set with the fermionic (4,4) \gls{CAS}. CAS(4,4) consists
of two pairs of orbitals which correlate at the dissociation
limit~\footnote{We should mention that for
  $\ce{d(O-H)} \le \SI{1.35}{\angstrom}$ the Hartree--Fock orbitals
  must be reordered.}. During \gls{QCC} iterations only the generators
with all-active indexes were ranked; four of them were included in the
\gls{QCC} ansatz for the next iteration. The convergence criterion was
the same as for \ce{N2}: the procedure was stopped when the absolute
maximum gradient \emph{in the active set} fell below 0.001.

\begin{figure}
  \centering %
  \includegraphics[width=1.0\linewidth]{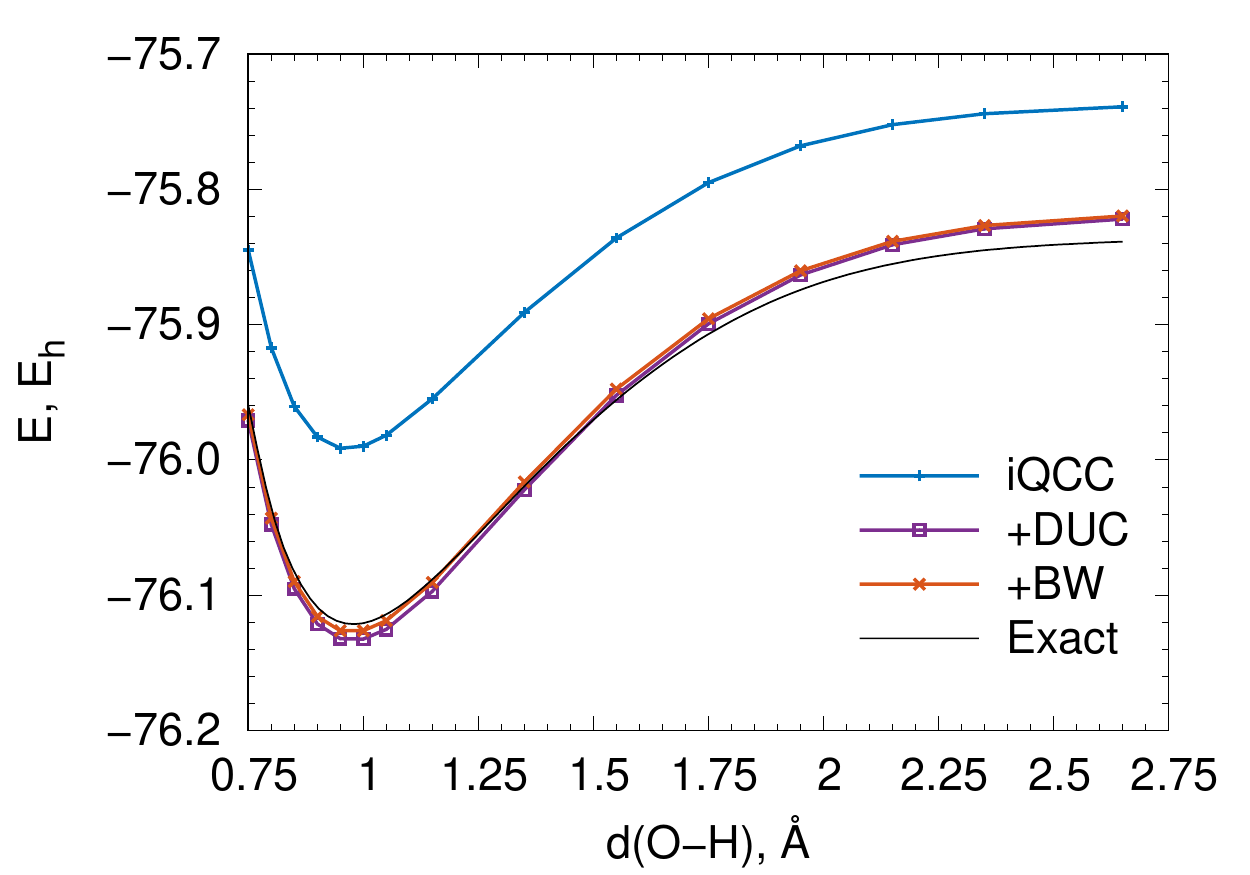}
  \caption{Potential energy curves for the symmetric \ce{H2O} molecule
    stretch. The active-space qubit treatment, equivalent to CAS(4,4)
    embedded into the \gls{FCI} (``exact'') problem (the Hartree--Fock
    orbital, which correlates with the 1s orbital of \ce{O} is frozen)
    in the 6-31G basis.}
  \label{fig:H2O_stretch}
\end{figure}

The resulting potential energy curves for the active-space \gls{iQCC}
method and its corrected counterparts are shown in
Fig.~\ref{fig:H2O_stretch}.
As can be seen, all the corrections to the \gls{iQCC} energies are
\emph{large} even at convergence, which is a hallmark of the
approximate nature of the bare active-space \gls{iQCC} energies. The
mean deviations from the exact curve decrease from approximately
\SI{117}{\milli\hartree} for the \gls{iQCC} curve to
\SI{2.1}{\milli\hartree} and \SI{5.8}{\milli\hartree} for DUC
[Eq.~\eqref{eq:DUC}] and BW [Eq.~\eqref{eq:SE_1}] corrections,
respectively; see also Table~\ref{tab:H2O_statistics}.
\begin{table}[!b]
  \centering
  \caption{Various integral characteristics of curves in
    Fig.~\ref{fig:H2O_stretch}. All quantities are in
    \si{\milli\hartree}.}
  \begin{minipage}{1.0\linewidth}
    \sisetup{table-format = 5.1, %
      round-mode = places, %
      round-precision = 0} %
    \begin{tabularx}{1.0\linewidth}{@{}lSSSS@{}}
      \toprule
      Method     & \multicolumn{3}{c}{Deviation, $(E - E_\text{exact})$} & \multicolumn{1}{r}{Non-parallelity}  \\
      \midrule
                 & \multicolumn{1}{c}{Max} & \multicolumn{1}{c}{Min}  & \multicolumn{1}{c}{Average}   & \\
      \cmidrule{2-4}
      iQCC       & 133.4673837  &  99.5898323 & 116.528608  & 33.8775514 \\
      iQCC + DUC &  16.6442714  & -12.442616  &   2.1008277 & 29.0868874 \\
      iQCC + BW  &  18.9141014  &  -7.2952664 &   5.8094175 & 26.2093678 \\
      \bottomrule
    \end{tabularx}
  \end{minipage}
  \label{tab:H2O_statistics}
\end{table}
While small mean deviations are important for some quantities, such as
singlet-triplet gaps, it is equally important to have curves that are
almost parallel to the exact one, to predict properties like
equilibrium geometries or vibrational frequencies, correctly. All the
corrections decrease the non-parallelity (the difference between
maximum and minimum deviations, see Table~\ref{tab:H2O_statistics})
for the whole considered region of \ce{O-H} distances, from 0.75 to
\SI{2.65}{\angstrom}. Moreover, both corrections drastically reduce
the non-parallelity error for the region near the equilibrium
geometry, $0.85 \le \ce{d(O-H)} \le \SI{1.15}{\angstrom}$: from
\SI{15}{\milli\hartree} to 4--\SI{5}{\milli\hartree}. This result is
consistent with the expectation that the perturbation theory-based
corrections are more reliable when correlation is not too strong.

\subsection{T1--S0 gap in the \ce{Ir(ppy)3} complex}
\label{sec:t1-s0-gap}

\begin{figure}
  \centering \includegraphics[width=0.60\linewidth]{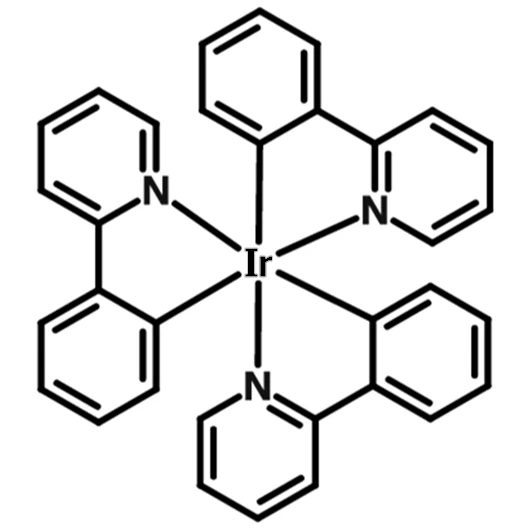}
  \caption{Tris-(2-phenylpyridine)iridium(III), \ce{Ir(ppy)3}.}
  \label{fig:Irppy3}
\end{figure}

Encouraged by performance of the perturbative corrections near
equilibrium configurations, we applied the enhanced \gls{iQCC} method
to a large, technologically relevant system, the
tris-(2-phenylpyridine)iridium(III), \ce{Ir(ppy)3} complex; see
Fig.~\ref{fig:Irppy3}.
This molecule is probably the most studied green phosphorescent
emitter~\cite{Baldo:1999/apl/4, Baldo:2000/nature/750,
  Holzer:2005/cp/93, Hofbeck:2010/ic/9290} widely used in organic
light-emitting diodes (OLEDs)~\cite{Sun:2006/nature/908}. Excited by
recombination of electrons and holes injected from the corresponding
transport layers, \ce{Ir(ppy)3} emits predominantly from the lowest
triplet state (the singlet state is also bright) with almost 100\%
quantum efficiency. For the targeted design of similar and more
advanced emitters it is important to reliably predict singlet-triplet
gaps (in either singlet and triplet equilibrium geometries), but due
to the size of molecules, the electronic-structure studies are
typically limited to the \gls{DFT}~\cite{Jansson:2007/cp/157,
  Wu:2008/jcp/214305, Smith:2011/cphc/2429}, with a prominent
counterexample~\cite{Kleinschmidt:2015/jcp/094301}.

The ground singlet electronic state \ce{Ir(ppy)3} is reasonably well
described by single-determinant Hartree--Fock or \gls{DFT} methods. On
the contrary, the triplet state is strongly multi-configurational with
a multitude of closely-spaced energy levels that arise from population
of low-lying d-orbitals of the Ir atom by an unpaired electron and its
delocalization across the conjugated system. If the unrestricted
Hartree--Fock method is used to describe the triplet state, the value
of $\braket{\hat S^2}$ is approximately 5, deviating markedly from the
ideal value of 2. Thus, the correct electronic structure is likely to
be determined by a balance of orbital mixing and correlation effects,
so that the \gls{CASSCF} method may be the best choice for the
problem; however, capturing dynamical correlation typically requires
unrealistically large active
spaces~\cite{vonBurg:2020/arXiv/2007.14460}. Thus, the \gls{iQCC}
method augmented with corrections could be the best-suited method for
such systems, since it accounts for the largest contributions
variationally, while considering the bulk of small ones perturbatively
allowing for the routine use of much larger active spaces.

To calculate the singlet-triplet gap we first optimized the molecular
structure of \ce{^3Ir(ppy)3} using the restricted open-shell
Hartree--Fock (ROHF) method, which guarantees the spin purity of a
Slater determinant. For the Ir atom we employed the 60-electron
relativistic effective core pseudopotential
(RECP)~\cite{Stevens:1992/cjc/612} with the partner double-zeta
valence basis set as implemented in \textsc{gamess}, and the Pople
split-valence double-zeta basis set augmented with polarization d
orbitals for the second-row atoms C and N, 6-31G(d). With
six-component d harmonics the atomic basis for the molecule contained
\num{622} functions. After the equilibrium geometry has been located,
\gls{CAS}(26e,~28o) has been created using 13 occupied and 15
unoccupied Hartree--Fock orbitals below and above the Fermi level,
respectively. The electronic Hamiltonian corresponding to that
\gls{CAS} was generated and converted into the 56-qubit operator using
the \gls{JW} mapping and a pairwise grouping ($1\alpha$, $1\beta$,
$2\alpha$, \textit{etc.}) of spin-orbitals; the resulting Hamiltonian
[Eq.~\eqref{eq:qubitH}] contained \num{901985} terms. A similar
procedure was carried out for the singlet state in the same geometry:
first, the restricted Hartree--Fock \glspl{MO} were generated, then
CAS(26e,~28o) was selected, and finally, the 56-qubit Hamiltonian was
assembled from the values of one- and two-electron integrals in the
chosen \gls{MO} basis using the \gls{JW} transformation resulting in
\num{901973}-term operator.

We were able to perform 10 \gls{iQCC} iterations using a
single-generator \gls{QCC} ansatz~\eqref{eq:qcc_form}, results are
shown in Table~\ref{tab:ST_Irppy3}.
\begin{table}
  \centering
  \caption{Various estimates for the singlet-triplet gap for
    \ce{Ir(ppy)3}.}
  \begin{minipage}{1.0\linewidth}
    \sisetup{table-format = 4.1, %
      round-mode = places, %
      round-precision = 2} %
    \begin{tabularx}{1.0\linewidth}{@{}lSS@{}}
      \toprule
      Method     &  \multicolumn{2}{c}{Singlet-triplet gap\footnotemark[1]} \\
      \midrule
                 & \multicolumn{1}{Y}{\si{\electronvolt}} & \multicolumn{1}{Y}{\si{\nano\meter}} \\
      \cmidrule{2-3}
      $\Delta$SCF(HF)           & 2.95  & 421 \\
      $\Delta$SCF(DFT/B3LYP\footnotemark[2])  & 2.26  & 549 \\
      $\Delta$MP2               & 2.58 & 481 \\[0.75ex]
      iQCC(1 iter) & \multicolumn{2}{c}{$= \Delta$SCF(HF)} \\
      iQCC(1 iter)  + DUC       & 2.57  & 483 \\[0.5ex]
      iQCC(5 iter)              & 2.99  & 415 \\
      iQCC(5 iter)  + DUC       & 2.56  & 485 \\[0.5ex]
      iQCC(10 iter)             & 2.98  & 416 \\
      iQCC(10 iter) + DUC       & 2.54  & 488 \\[0.75ex]
      Exp.\footnotemark[3]      & 2.52  & 491 \\
      \bottomrule
    \end{tabularx}
    \footnotetext[1]{In the triplet geometry optimized at the
      restricted open-shell Hartree--Fock level.} %
    \footnotetext[2]{Refs.~\citenum{Becke:1993/jcp/5648} and
      \citenum{Stephens:1994/jpc/11623}.} %
    \footnotetext[3]{In $\sim 10^{-5}$\,M solution in 2-MeTHF at
      \SI{77}{\kelvin}; see Ref.~\citenum{Sajoto:2009/jacs/9813}.} %
  \end{minipage}
  \label{tab:ST_Irppy3}
\end{table}
It appears that the most of the correction to the $\Delta$SCF can be
captured perturbatively, as both $\Delta$MP2 and ``iQCC(1 iter) +
DUC'' results are already quite close to the reference value. The bare
\gls{iQCC} estimates behave non-monotonically, which is explained by
slightly different rates of convergence of the absolute \gls{iQCC}
energies for singlet and triplet states. Corrected \gls{iQCC} values
(``+ DUC''), however, exhibit the monotonic convergence. Overall, the
\gls{iQCC} procedure converges slowly, and without perturbative
corrections such a situation is detrimental for the \gls{iQCC} method,
as one would need to carry out prohibitively many iterations to reach
the desired accuracy.

\section{Conclusions}
\label{sec:conclusions}

We have developed and tested several numerical techniques that are
aimed to \textit{a posteriori} correct energies computed by the
\gls{iQCC} method. They are rooted in a new representation for qubit
Hamiltonians, Eq.~\eqref{eq:Ising_decomp}, in which regular and simple
rules exist for evaluating matrix elements with the direct-product
qubit states, Eqs.~\eqref{eq:omega_defs} and \eqref{eq:X_av_pmf}. In
this respect it has a lot in common with normal-ordered fermionic
Hamiltonians. The new decomposition naturally leads to a linear
variational ansatz~\eqref{eq:linear_variational_ansatz}. Being not
suitable for quantum computers, it allowed us to formulate the
\gls{CI}-like variational problem~\eqref{eq:CI_problem} and the qubit
form of the second-order \acrlong{ENPT}, Eq.~\eqref{eq:ENPT2_energy};
both can be efficiently evaluated on a \emph{classical} computer.

Two essential modifications for the aforementioned techniques were
also developed. The first is aimed to circumvent the divergence of the
\gls{ENPT} series due to small denominators in the strong correlation
regime [see Eq.~\eqref{eq:DUC}]. The second is the unified
variational-perturbational scheme, which ``interpolates'' between
purely variational and perturbative solutions [see
Sec.~\ref{sec:comb-vari-pert}] for flexible control of computational
efforts. The unified scheme in the limit of a trivial $1\times 1$
effective Hamiltonian matrix reduces to the second-order
Brillouin--Wigner perturbation theory.

Operationally, the new perturbative corrections require computing of a
few extra quantities, namely, the mean-field excited energies
[Eq.~\eqref{eq:E_k}]; everything else is available as elements of the
original \gls{iQCC} procedure. All these steps are not the
computational bottleneck even for the largest problem considered here
and can be safely offloaded to a classical computer. By capturing some
of energy contributions classically, we additionally decrease the use
of quantum resources. First of all, the total number of \gls{iQCC}
iterations is decreased for the same accuracy requirements. Secondly,
by limiting the qubit indices of generators to be in the active set,
we limit the growth of intermediate Hamiltonians and, thus, the number
of quantum measurements needed at subsequent iterations.

Numerical assessment of the proposed techniques on several
prototypical problems, namely, 10-qubit \ce{N2} dissociation, the
24-qubit symmetric water molecule stretch, and finally, and
large-scale, 56-qubit simulations of the singlet-triplet gap in the
\ce{Ir(ppy)3} complex have demonstrated a substantial improvement over
the original \gls{iQCC} method. New corrections address the most
severe shortcoming of the original \gls{iQCC} method: its numerical
inefficiency in the case of weak correlation; they also provide
additional flexibility in cases when only a subset of qubits is
strongly correlated.


\bibliography{encorr}

\end{document}